\documentclass[aps,prb,superscriptaddress,amsmath,amssymb,showpacs,floatfix,reprint,longbibliography]{revtex4-2}

\usepackage{graphicx}
\usepackage{float}

\usepackage{multirow}    
    
\usepackage{color}
\usepackage[colorlinks=true, urlcolor=blue, linkcolor=blue, citecolor=blue, pdftex]{hyperref}

\definecolor{karlocolor}{rgb}{0.9,0.2,0}
\definecolor{oldcolor}{rgb}{0,0.3,0.6}
\definecolor{newcolor}{rgb}{0,0.6,0.3}

\begin{document}

\title{The dynamical structure factor of the SU(4) algebraic spin liquid on the honeycomb lattice}

\author{D\'aniel V\"or\"os}
\affiliation{Department of Theoretical Physics, Institute of Physics,
Budapest University of Technology and Economics, Műegyetem rakpart 3, H-1111 Budapest, Hungary}
\affiliation{Institute for Solid State Physics and Optics, HUN-REN Wigner Research Centre for Physics, H-1525 Budapest, P.O. Box 49, Hungary}
\author{Karlo Penc}
\affiliation{Institute for Solid State Physics and Optics, HUN-REN Wigner Research Centre for Physics, H-1525 Budapest, P.O. Box 49, Hungary}

\date{\today}

\begin{abstract}
We compute the momentum resolved dynamical spin structure factor $S(k,\omega)$ of the SU(4) Heisenberg model on the honeycomb lattice assuming the $\pi$-flux Dirac spin liquid ground state by two methods: (i) variationally using Gutzwiller projected particle-hole excitations of the $\pi$-flux Fermi sea and (ii) in the non-interacting parton mean-field picture. 
The two approaches produce qualitatively similar results. Based on this analogy, we argue that the energy spectrum of the projected excitations is a gapless continuum of fractional excitations. 
Quantitatively, the Gutzwiller projection shifts the weight from higher to lower energies, thus emphasizing the lower edge of the continuum. In the mean-field approach, we obtained the $1/\text{distance}^4$ decay of the spin correlation function, and the local correlations show $S^{33}_{\text{MF}}(\omega)\propto \omega^3$ behavior.
\end{abstract}


\maketitle


\section{Introduction}

Early on, Li {\it et al.}  pointed out that the highest symmetry that transition metal compounds with $S=1/2$ spins and degenerate $e_g$ orbitals per site can have is SU(4) \cite{PhysRevLett.81.3527}. Then, they can be described by the SU(4) symmetric Heisenberg model,
\begin{equation}
\label{eq:Heisenberg_Hamiltonian}
    \mathcal{H} = J \sum_{\langle i, j\rangle} \sum_{a=1}^{15} T_{i}^{a} T_{j}^{a},
\end{equation}
where the summation is over the $\langle i, j\rangle$ nearest neighbor sites and $T_{i}^{a}$, with $a=1,2,\dots 15$, are the generators of the SU(4) Lie algebra in the four-dimensional fundamental (also known as defining) representation.
Realizing such a high symmetry in real materials is not guaranteed. 
It is suggested that the spin-orbital interactions for face-sharing MO$_6$ octahedra in  \cite{PhysRevB.91.155125} and the $J_{\text{eff}} = 3/2$ quartets in the $d^1$ electron configurations in the strong spin-orbit coupling limit \cite{2018PhRvL.121i7201Y,*PhysRevB.104.224436} may lead to SU(4) symmetry. 
Fermionic cold atoms in optical lattices also provide promising systems by tailoring the exchange between $S=3/2$ atoms \cite{PhysRevLett.95.266404}, and with alkaline earth atoms \cite{PhysRevLett.103.135301, Gorshkov:2010aa, 2018PhRvL.121v5303O}. 
Recently, the SU(4) Heisenberg model has been proposed to describe the low energy properties of magic-angle twisted graphene \cite{PhysRevB.98.045103, PhysRevB.98.245103} and metal dichalcogenide bilayers 
\cite{Angeli2021, Xian:2021aa, PhysRevLett.127.247701}.

The ground state of the SU(4) Heisenberg model depends on the underlying lattice. The one-dimensional SU(4) Heisenberg chain is critical, with algebraically decaying spin correlations \cite{PhysRevB.58.9114, 1999PhRvL..82..835F}, although the phase diagram of one-dimensional alkaline-earth atoms is more complex \cite{CAPPONI201650}. In coupled chains (ladders), singlet plaquettes made of four spins appear, a feature typical of SU(4) spins \cite{PhysRevLett.86.4124, PhysRevB.72.214428, PhysRevB.74.224426, PhysRevB.98.085104}. The ground state is dimerized on the square lattice and breaks the SU(4) symmetry  \cite{PhysRevLett.107.215301}. 
The model's fate on the triangular lattice, initially considered for LiNiO$_2$ in Ref.~\onlinecite{PhysRevB.70.014428}, is still open. Refs.~\onlinecite{PhysRevB.68.012408,PhysRevLett.127.247701,PhysRevResearch.3.023138} suggest the importance of the SU(4) singlet plaquettes, but a gapless uniform \cite{PhysRevLett.125.117202} or stripy \cite{PhysRevResearch.3.023138, JIN2022918} spin liquid phase, and a trimerized state breaking the flavor symmetry \cite{Anna_trimerization} are also possible candidates.

\begin{figure}
\includegraphics[width=0.95\columnwidth]{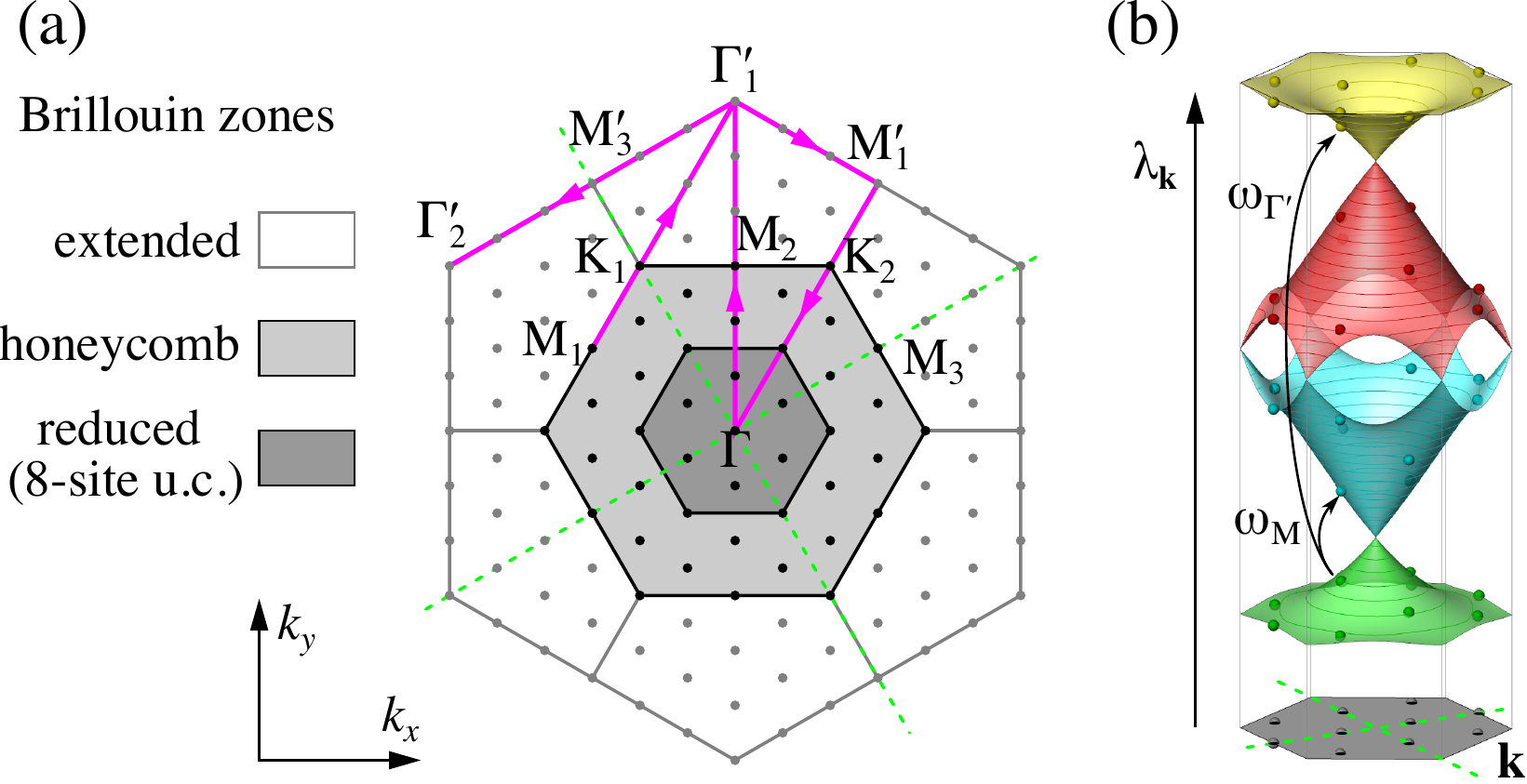}
\caption{
(a) The reciprocal space of the honeycomb lattice. The reduced Brillouin zone (a quarter of the honeycomb lattice Brillouin zone) originates from the 8-site unit cell of the $\pi$-flux $\mathcal{H}_{\text{MF}}$. The structure factor is fully characterized by its behavior in the extended zone. The momenta of the points $K_1$ and $K_2$  are $(\pm 2\pi/3,2\pi/\sqrt{3})$. The dots are the $k$-points of the $N_s=72$ site cluster; we plot the structure factor along the magenta path.
(b) The twofold degenerate bands of the parton mean-field theory, defined by Eq.~(\ref{eq:characteristic_polynomial}), in the reduced Brillouin zone, with Dirac cones at $\lambda=\pm \sqrt{3} t$ and $\mathbf{k}=0$. The tiny spheres represent the $(\lambda_{\mathbf{k}},\mathbf{k})$ points for the fermions on the 72-site cluster with antiperiodic boundaries. Instead of the $D_6$ symmetry, only the $D_2$ remains; the green dashed lines show the invariant reflections. The arrows indicate the lowest energy particle-hole excitations with nonvanishing weight at the $\text{M}$ and $\Gamma'$ momenta.  
\label{fig:BZ_and_bands}
}
\end{figure}

 On the honeycomb lattice, exact diagonalization of small clusters and iPEPS calculations found no evidence for symmetry breaking, and a Gutzwiller projected $\pi$-flux parton Fermi sea was proposed as the ground state \cite{SU4_Honeycomb}. This was also confirmed by parton mean-field \cite{Jakab_PhysRevB.93.064434_2016} and DMRG \cite{PhysRevB.107.L180401} studies and it was found to be robust to small anisotropies \cite{PhysRevB.100.205131}. However, a finite temperature DMRG calculation revealed evidence for gap opening \cite{PhysRevB.105.L201115, Masahiko_private}.

\begin{figure}[tb]
\centering
\includegraphics[width=0.95\columnwidth]{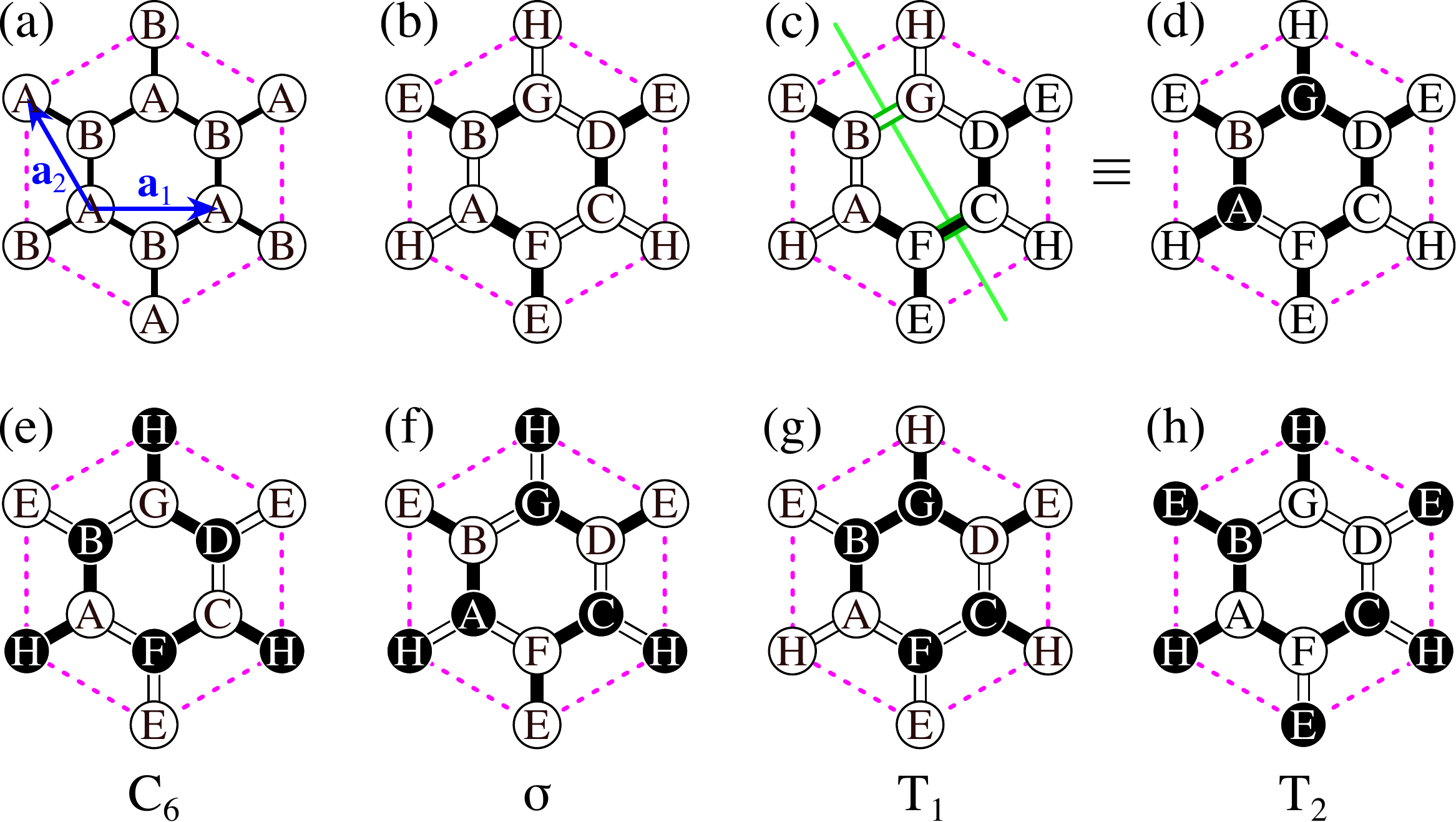}

\caption{
\label{fig:lattice}
(a) A $D_6$ symmetric eight-site cluster of the honeycomb lattice with periodic boundary conditions, containing four unit cells, each with a basis of two sites (A and B). 
The arrows show the primitive vectors $\mathbf{a}_1$ and $\mathbf{a}_2$.
(b) The minimal, eight-site cluster of  $\pi$-flux $\mathcal{H}_{\text{MF}}$ with a basis A, B, \dots H, invariant under a $\mathsf{C}_3$ rotation and translations by $2 \mathbf{a}_1$ and $2 \mathbf{a}_2$.
The black and white bonds represent hopping amplitudes of opposite signs; their product around a plaquette is negative, $-|t|^6$. 
(c) Antiperiodic boundary condition reverses the signs of the hoppings crossing the cluster's boundary, shown with the green line. 
(d) The hoppings with periodic boundary conditions used in Ref.~\onlinecite{SU4_Honeycomb}. They are gauge equivalent to those on (c) if one reverses the sign of the fermionic operators on sublattices A and G (black circles). For certain cluster sizes, the equivalence requires antiperiodic boundary conditions.
(e)  The hopping amplitudes after a $\mathsf{C}_6$ rotation around the center of the middle hexagon, (f) reflection about the vertical axis, and
(g) and (h)  translations by $\mathbf{a}_1$ and $\mathbf{a}_2$.
For periodic boundaries, gauge transformations where fermionic operators on sublattices marked with black circles get multiplied by a minus sign restore the original hopping structure (b).
}
\end{figure}

\section{Mean-field ground state}

In the fermionic parton mean field \cite{Affleck_pi_flux, *PhysRevB.39.11538}, the SU(4) spin operators in Eq.~(\ref{eq:Heisenberg_Hamiltonian}) are decomposed into fermions,
\begin{equation}
  T_i^{a} = \frac{1}{2} \sum_{\alpha, \beta = 1}^{4} c^{\dagger}_{i, \alpha} \lambda^{a}_{\alpha, \beta} c^{\phantom{\dagger}}_{i, \beta} \;
  \label{eq:spin_operator_with_partons_SU_4}
\end{equation}
where the  $4\times4$ matrices $\lambda^{a}$ generalize the eight SU(3) Gell-Mann matrices \cite{Gell_Mann_matrices} to SU(4). In the fundamental representation, a singly occupied site with a fermion of flavor $\alpha \in \{1,2,3,4\}$ represents a spin basis state $\alpha$, and the system is quarter filled.
We obtain the mean-field approximation by substituting Eq.~(\ref{eq:spin_operator_with_partons_SU_4}) into the Hamiltonian (\ref{eq:Heisenberg_Hamiltonian}) and replacing a pair of fermionic operators with their expectation value. 
We consider the time-reversal invariant ansatz \cite{SU4_Honeycomb}
\begin{equation}
\label{eq:P_G_pi_flux_FS_mean_field_Hamiltonian}
\mathcal{H}_{\text{MF}} = \sum_{\alpha = 1}^{4} \sum_{\langle i,j \rangle} t_{i,j} c^{\dagger}_{i,\alpha} c^{\phantom{\dagger}}_{j,\alpha},
\end{equation}
where each hopping amplitude is real with $t_{i,j}= \pm |t|$ and the signs of the hoppings are shown in Fig.~\ref{fig:lattice}(b). The energy of the quarter-filled honeycomb lattice is minimal for the $\pi$-flux state \cite{Hasegawa_PhysRevLett.63.907_1989}, when the product of the hopping amplitudes around any plaquette is negative. 
In the reciprocal space, the $\mathcal{H}_{\text{MF}}$ is an eight-by-eight matrix with the characteristic polynomial
\begin{equation}
0=  \left( \lambda_{\mathbf{k}}^4 - 6 t^2 \lambda_{\mathbf{k}}^2 + t^4 \gamma_{\mathbf{k}} \right)^2 ,
\label{eq:characteristic_polynomial}
\end{equation}
where 
\begin{equation}
\gamma_{\mathbf{k}} = 3  + 2 \cos 2\mathbf{k} \cdot \mathbf{a_1}  + 
2 \cos 2\mathbf{k} \cdot \mathbf{a_2} + 2 \cos 2\mathbf{k} \cdot [\mathbf{a_1} + \mathbf{a_2} ] .
\end{equation} 
The one-particle spectrum $\lambda_{\mathbf{k}}$ consists of four bands, each doubly degenerate, as shown in Fig.~\ref{fig:BZ_and_bands}(b). 
The quarter-filled $\pi$-flux Fermi sea  $|\pi \text{FS}\rangle $ fills the lowest band up to the Dirac Fermi point at $\mathbf{k}=0$. 
Solving the 
$t_{i,j} = J \sum_{\beta} \langle \pi \text{FS}|  c^{\dagger}_{i,\beta} c^{\phantom{\dagger}}_{j,\beta} | \pi \text{FS} \rangle$ 
self-consistency equation, we get 
$t = (0.7712 \pm 0.0002) J$ in the thermodynamic limit, 
in agreement with Ref.~\onlinecite{Jakab_PhysRevB.93.064434_2016}.  
The $|\pi \text{FS}\rangle$ also contains charge fluctuations, i.e., sites with multiple occupancies that do not map to a spin state.
To go beyond the mean-field approximation, one enforces single occupancy by applying the Gutzwiller projector $P_{\text{G}}$ to $|\pi \text{FS}\rangle$. 
%
%
The projective symmetry group classification on the honeycomb lattice and its stability analysis for the SU(2) case found that the Dirac spin liquid is an unstable ground state \cite{Song_NatCom_2019,Lu_PRB_2011}. For SU(4), the fluctuations around the mean-field solution are weaker, so the ansatz may become stable (the possible instabilities of the $\pi$-flux state on the honeycomb lattice are discussed in Refs. \onlinecite{calvera2021theory,mondal2023emergent}).

Here we extend the study of the SU(4) Heisenberg model on the honeycomb lattice by including Gutzwiller projected particle-hole excitations above the $\pi$-flux Fermi sea. 
These states represent fractionalization within the fermionic parton construction and have been used in \cite{First_S_q_w_latter,*Yang_Li_2011PhRvB..83f4524Y, Excited_states_above_Fermi_sea, Mei_Wen_arxiv_2015,2020PhRvB.102s5106Z, Becca, 2018PhRvB..98m4410Y,2020PhRvB.102a4417F, 2019PhRvX...9c1026F} to describe the dynamics of SU(2) quantum spin liquids. In \cite{PhysRevB.104.184426}, we have shown that projected particle-hole excitations capture the main features of the dynamical structure factor of the SU(3) Heisenberg chain. 
Below, we implement the method to the SU(4) honeycomb model, calculate the dynamical structure factor, and argue that the spectrum is gapless at the $\text{M}$ and $\text{M}'$ points.

Before proceeding to the actual calculation, let us briefly review the projective symmetries of the $\pi$-flux ansatz.
The basis of the honeycomb lattice consists of two sites, with basis vectors 
$\boldsymbol{\delta}_{A} = (0,0)$ 
and 
$\boldsymbol{\delta}_{B} = (0,\frac{1}{\sqrt{3}})$. The position of site $j$ is 
 $\mathbf{r}_j = \mathbf{R}_j +  \boldsymbol{\delta}_{d_j}$, where $\mathbf{R}_j = R_{j1} \mathbf{a}_1 + R_{j2} \mathbf{a}_2$ is a Bravais lattice vector and $d_j \in \{ A,B \}$.
The $\mathsf{C}_6$ sixfold rotation, the $\sigma$ reflection (generating the $D_6$ point group), and the elementary translations $\mathsf{T}_1$ and $\mathsf{T}_2$ by the primitive vectors $\mathbf{a}_1 = (1,0)$ and $\mathbf{a}_2 = (-\frac{1}{2},\frac{\sqrt{3}}{2})$ shown in Fig.~\ref{fig:lattice}(a) generate the $p6m$ wallpaper group of the lattice. 
However, the $\pi$-flux $\mathcal{H}_{\text{MF}}$ (\ref{eq:P_G_pi_flux_FS_mean_field_Hamiltonian}) and $|\pi \text{FS} \rangle$ break some of these symmetries. The hopping structure shown in Fig.~\ref{fig:lattice}(b) requires a quadrupled unit cell containing a basis of eight sites, with a remaining $C_3$ point group.
In addition to geometrical symmetries, we can apply gauge transformations to the fermions. 
A site-dependent, but flavor-independent transformation $G:c^{{\dagger}}_{j,\alpha} \rightarrow e^{i \phi(j)}c^{{\dagger}}_{j,\alpha}$  leaves the spin operators in Eq.~(\ref{eq:spin_operator_with_partons_SU_4}) unchanged.
One can use this gauge redundancy to restore the symmetries of $\mathcal{H}_{\text{MF}}$, if the symmetry operations $\mathsf{g} \in \lbrace \mathsf{C}_6, \sigma, \mathsf{T}_1, \mathsf{T}_2 \rbrace$ are combined with an appropriate gauge transformations $G_{\mathsf{g}} \in  \lbrace G_{\mathsf{C}_6}, G_{\sigma}, G_{\mathsf{T}_{1}}, G_{\mathsf{T}_{_2}} \rbrace $ 
into
$\tilde{\mathsf{g}} \equiv G_{\mathsf{g}}\mathsf{g}$, 
so that
\begin{equation}
\mathcal{H}_{\text{MF}} = G_{\mathsf{g}} \mathsf{g} \mathcal{H}_{\text{MF}} \mathsf{g}^{-1} G^{-1}_{\mathsf{g}}= \tilde{\mathsf{g}} \mathcal{H}_{\text{MF}} \tilde{\mathsf{g}}^{-1}.
\label{eq:projective_symmetry_of_H_MF} 
\end{equation}
This is the projective symmetry introduced in Refs. \onlinecite{First_Parton_construction_article_Jain, *Second_Parton_construction_article, Projective_construction_Wen, *Projective_Symmetry_Group_Wen}. 
The existence of a gauge transformation $G_{\mathsf{g}}$, for which $\mathcal{H}_{\text{MF}}$ has projective symmetry for $\tilde{\mathsf{g}}$, guarantees that the expectation values of spin operators and their correlation functions are symmetric under $\mathsf{g}$.

The condition imposed by Eq.~(\ref{eq:projective_symmetry_of_H_MF}) is equivalent to 
\begin{equation}
t_{i,j}  =e^{i \phi_{\mathsf{g}}(i)}  t_{\mathsf{g}^{-1}(i),\mathsf{g}^{-1}(j)} e^{-i\phi_{\mathsf{g}}(j)},
\label{eq:projective_symmetry_for_hoppings}
\end{equation}
where the fermionic operators transform  as
\begin{equation}
    \tilde{\mathsf{g}} c^{{\dagger}}_{m,\alpha}\tilde{\mathsf{g}}^{-1} = e^{i \phi_{\mathsf{g}}(\mathsf{g}(m))} c^{{\dagger}}_{\mathsf{g}(m),\alpha}.
\label{eq:gauge_transformation_of_a_fermionic_creation_operator}
\end{equation}
Figs.~\ref{fig:lattice}(e)-(h) illustrate the effect of the symmetry operations on the hopping amplitudes $t_{m,n} c^{{\dagger}}_{m,\alpha} c^{\phantom{\dagger}}_{n,\alpha} \rightarrow t_{m,n} c^{{\dagger}}_{\mathsf{g}(m),\alpha} c^{\phantom{\dagger}}_{\mathsf{g}(n),\alpha}$ drawn in Fig.~\ref{fig:lattice}(b). 
The sign changes following a symmetry operation can be reversed by multiplying each fermionic operator on the sublattices marked with black circles by minus signs as $c^{{\dagger}}_{j,\alpha} \to -c^{{\dagger}}_{j,\alpha}$ and $c^{\phantom{\dagger}}_{j,\alpha} \to -c^{\phantom{\dagger}}_{j,\alpha}$. These minus signs can then be transferred to the hopping amplitudes $t_{i,j}$ connected to the site $j$. 
These are the gauge transformations we are looking for, so $\mathcal{H}_{\text{MF}}$ of Eq.~(\ref{eq:P_G_pi_flux_FS_mean_field_Hamiltonian}) has projective symmetry for all generators $\lbrace \mathsf{C}_6, \sigma, \mathsf{T}_1, \mathsf{T}_2 \rbrace$. 
The sign patterns are periodic in $2 \mathbf{a}_1$ and $2 \mathbf{a}_2$ and fit into the quadrupled unit cell of Fig.~\ref{fig:lattice}(b). 
For translations, the signs can be written as 
$e^{i \phi_{\mathsf{T}_{1}}(j)} = e^{i \mathbf{Q}_1 \cdot \mathbf{r}_j }$
and $e^{i \phi_{\mathsf{T}_2}(j)} = e^{i \mathbf{Q}_2 \cdot \mathbf{r}_j }$, 
where 
$\mathbf{Q}_1 = ( \pi , -\sqrt{3}\pi)$ 
and 
$\mathbf{Q}_2 = ( \pi , \sqrt{3}\pi)$ 
are the momenta of the $\text{M}'$ points at the edge of the extended Brillouin zone, see Fig.~\ref{fig:BZ_and_bands}. 
%
The gauge transformation of any symmetry operation can be found by applying the generators $\{ \mathsf{C}_6, \sigma, \mathsf{T}_1, \mathsf{T}_2\}$ one after the other, using that for a product $\mathsf{g}_1 \mathsf{g}_2$ of the $\mathsf{g}_1$ and $\mathsf{g}_2$ symmetry operations
%
\begin{align}
    \tilde{\mathsf{g}}_2 
    (\tilde{\mathsf{g}}_1  c^{{\dagger}}_{m,\alpha}\tilde{\mathsf{g}}^{-1}_1)\tilde{\mathsf{g}}^{-1}_2 
 & = e^{i \phi_{\mathsf{g}_2}(\mathsf{g}_2\mathsf{g}_1(m))} e^{i \phi_{\mathsf{g}_1}(\mathsf{g}_1(m))} c^{{\dagger}}_{\mathsf{g}_2\mathsf{g}_1(m),\alpha} \nonumber \\
    & = e^{i \phi_{\mathsf{g}_2\mathsf{g}_1}(\mathsf{g}_2\mathsf{g}_1(m))}c^{{\dagger}}_{\mathsf{g}_2\mathsf{g}_1(m),\alpha}.
\label{eq:gauge_transformation_of_a_combined_symmetry}
\end{align}

Unfortunately, the $|\pi \text{FS} \rangle$ is degenerate for periodic boundary conditions. 
Imposing antiperiodic boundaries, as shown in Fig.~\ref{fig:lattice}(c), makes the variational ground state unique, although it breaks the projective symmetries for the sixfold rotation and some of the reflections (see Appendix \ref{sec:appendix_APBC} for details).
Consequently, the spin correlation function $\langle \pi \text{FS}| T^3_{\mathbf{R},d} T^3_{\mathbf{R'},\bar{d}} |\pi \text{FS} \rangle$ is symmetric only under translations and the $D_2$ point group, which is also confirmed by numerical results.
Since the asymmetry is due to the boundary conditions, we expect the restoration of the $D_6$ symmetry in the thermodynamic limit, in agreement with the numerics.

\section{Dynamical structure factor}

The momentum-resolved dynamical spin structure factor at zero temperature is defined by 
\begin{equation}
\label{eq:S^33_k_omega}
     S^{33}(\mathbf{k},\omega) =  \sum_{f} \left| \langle f| T^{3}_{\mathbf{k}}| \text{GS} \rangle  \right|^2
     \delta(\omega + E_{\text{GS}} - E_{f})\;,
\end{equation}
where the sum is over the final states $|f\rangle$ of $\mathcal{H}$ with energies $E_{f}$, $|\text{GS}\rangle$ denotes the ground state with energy $E_{\text{GS}}$, and 
$T^{3}_{\mathbf{k}} = \frac{1}{\sqrt{N_s}} \sum_{\mathbf{R},d} e^{i \mathbf{k} \cdot (\mathbf{R}+\boldsymbol{\delta}_d)} T^{3}_{\mathbf{R},d}$ is the Fourier transform of the 
\begin{equation}
 T^{3}_{\mathbf{R},d} = \frac{1}{2} \left( c^{\dagger}_{\mathbf{R},d,1} c^{\phantom{\dagger}}_{\mathbf{R},d,1} - c^{\dagger}_{\mathbf{R},d,2} c^{\phantom{\dagger}}_{\mathbf{R},d,2} \right)
 \label{eq:T3def}
\end{equation}
diagonal spin operator on the sublattice $d$, as defined in Eq.~(\ref{eq:spin_operator_with_partons_SU_4}), and $N_s$ denotes the number of lattice sites in the cluster. 
%
%
$S^{33}(\mathbf{k},\omega)$ satisfies the sum rule \ref{sec:appendix_sum_rule}
\begin{equation}
 \sum_{\mathbf{k}\in \text{eBZ}}\int S^{33}(\mathbf{k},\omega) \text{d} \omega = \frac{3}{16}N_{s},
 \label{eq:sum_rule_GP}
\end{equation}
where the sum is over all $\mathbf{k}$ vectors in the extended Brillouin zone (eBZ) shown in Fig.~\ref{fig:BZ_and_bands}(a).
%

%
%

Following Refs.~\onlinecite{First_S_q_w_latter,*Yang_Li_2011PhRvB..83f4524Y, Excited_states_above_Fermi_sea, Becca}, we construct an approximation to the states $|f \rangle$ by diagonalizing the Hamiltonian in a Hilbert subspace spanned by Gutzwiller-projected  particle-hole states with momenta $\mathbf{k}$, 
\begin{align}
|\mathbf{k};\alpha; \mathbf{R}, d;\bar{d}\rangle &= P_{\text{G}} \sum_{\mathbf{R'}} e^{i \mathbf{k} \cdot \mathbf{R'}} 
\mathsf{\tilde{T}}_1^{R'_1}  
\mathsf{\tilde{T}}_2^{R'_2}  
c^{{\dagger}}_{\mathbf{R},d,\alpha}c^{\phantom{\dagger}}_{0,\bar{d},\alpha} | \pi \text{FS} \rangle 
\nonumber \\
=P_{\text{G}} &\sum_{\mathbf{R'}} e^{i \mathbf{k} \cdot \mathbf{R'}}
e^{i (R_1' \mathbf{Q}_1 + R_2' \mathbf{Q}_2) \cdot (\mathbf{R} + \boldsymbol{\delta}_{d} + \boldsymbol{\delta}_{\bar{d}})} \nonumber \\  
&\times (-1)^{\xi (\mathbf{R},\mathbf{R'})} c^{{\dagger}}_{\mathbf{R+R'},d,\alpha} c^{\phantom{\dagger}}_{\mathbf{R'},\bar{d},\alpha} |\pi\text{FS} \rangle ,
\label{eq:particle-hole_excited_states}
\end{align}
where
$\mathbf{R'} = R'_1 \mathbf{a_1} + R'_2 \mathbf{a_2}$ and $d, \bar{d} \in \{A, B\}$. The $e^{i (R_1' \mathbf{Q}_1 + R_2' \mathbf{Q}_2) \cdot (\mathbf{R} + \boldsymbol{\delta}_{d} + \boldsymbol{\delta}_{\bar{d}})}(-1)^{\xi (\mathbf{R},\mathbf{R'})}$ comes from
$\tilde{\mathsf{T}}_1^{R'_1} \tilde{\mathsf{T}}_2^{R'_2}$ 
using Eq.~(\ref{eq:gauge_transformation_of_a_combined_symmetry}),  $e^{i \mathbf{Q}_{1,2} \cdot {2 \mathbf{a}_{1,2}}} = 1$, and $\mathsf{\tilde{T}}_l |\pi \text{FS} \rangle = |\pi \text{FS} \rangle$. The $(-1)^{\xi (\mathbf{R},\mathbf{R'})}$ accounts for the boundary conditions, it is always $+1$ for periodic boundaries, while for antiperiodic boundaries we get $-1$ when $\mathbf{R}$ is inside the cluster, but $\mathbf{R+R'}$ crosses the antiperiodic boundaries an odd number of times (it comes from the $G^{\text{APBC}}_{\mathsf{T}_{l}}$ hidden in $\tilde{\mathsf{T}_{l}}$, see Appendix~\ref{sec:appendix_APBC}). 
The states 
\begin{equation}
|\mathbf{k};\mathbf{15}_3; \mathbf{R}, d;\bar{d}\rangle \equiv
 \frac{1}{2} \left( |\mathbf{k};1; \mathbf{R}, d;\bar{d}\rangle - |\mathbf{k};2; \mathbf{R}, d;\bar{d}\rangle \right)
\end{equation}
belong to the 15-dimensional adjoint irreducible representation of SU(4), while the $\sum_{\alpha = 1}^4 |\mathbf{k};\alpha; \mathbf{R}, d;\bar{d}\rangle$ are singlets.
We calculate the overlap 
$\langle \mathbf{k}; \mathbf{15}_3; \mathbf{R}, d;\bar{d} | \mathbf{k}; \mathbf{15}_3; \mathbf{R'}, d';\bar{d}'  \rangle$ 
and the Hamiltonian matrix 
$\langle \mathbf{k}; \mathbf{15}_3; \mathbf{R}, d;\bar{d} | \mathcal{H}|\mathbf{k}; \mathbf{15}_3; \mathbf{R'}, d';\bar{d}'  \rangle$ by Monte Carlo sampling. 
Solving the generalized eigenvalue problem provides the excitation energies $E_f$ and the states $|f\rangle$, from which we can calculate $\langle f | T^3_{\mathbf{k}} | \text{GS} \rangle =  \sum_{d} e^{i \mathbf{k} \cdot \delta_d}\langle f | \mathbf{k}; \mathbf{15}_3; \mathbf{0}, d;d  \rangle $, as Eq.~(\ref{eq:T3def}) implies. 
Repeating the same using the singlet states we get the singlet excitation energies.
For details, see Ref.~\onlinecite{PhysRevB.104.184426}.

\begin{figure}[h!]
\includegraphics[width=0.95\columnwidth]{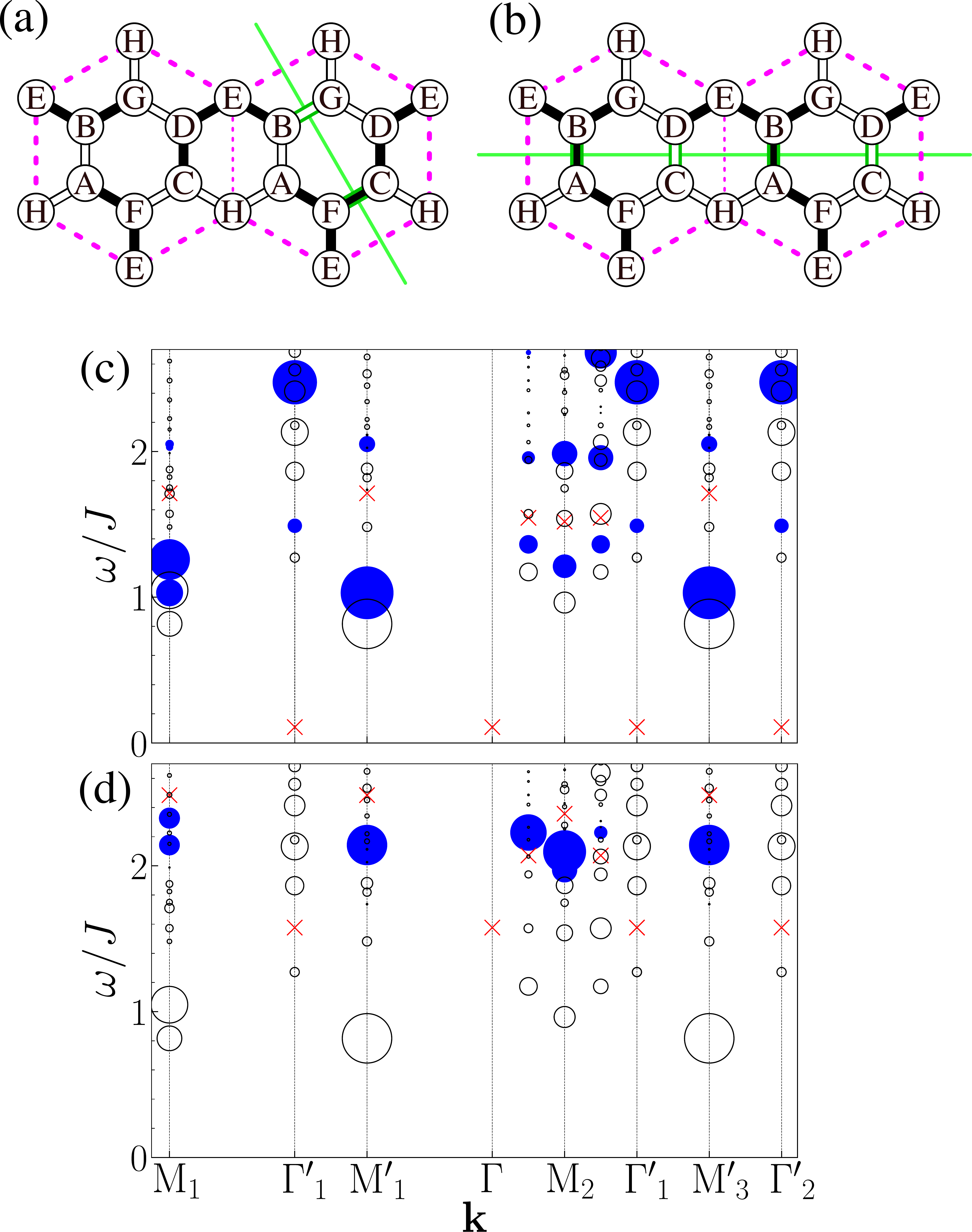}
\caption{(a) and (b) the $N_s=16$ site clusters with different antiperiodic boundaries (dashed green lines) which change the sign of the hoppings they cross. 
 (c) The $S^{33}(\mathbf{k},\omega)$ for the 16 site clusters in (a), calculated by variational method (blue circles) and by exact diagonalization (black circles). The red crosses indicate the lowest singlets, from which we can see that the variational ground state energy of the $|\pi \text{FS}\rangle$ state is close to the ED ground state energy, which is at $\omega = 0$.
 (d) The $S^{33}(\mathbf{k},\omega)$ for the 16 site clusters in (b), in this case, the agreement is not so good. We expect the sensitivity to the boundary conditions to decrease with larger clusters.}
\label{fig:S_q_omega_16}
\end{figure}

\begin{figure}[t]
\centering
\includegraphics[width=0.48\textwidth]{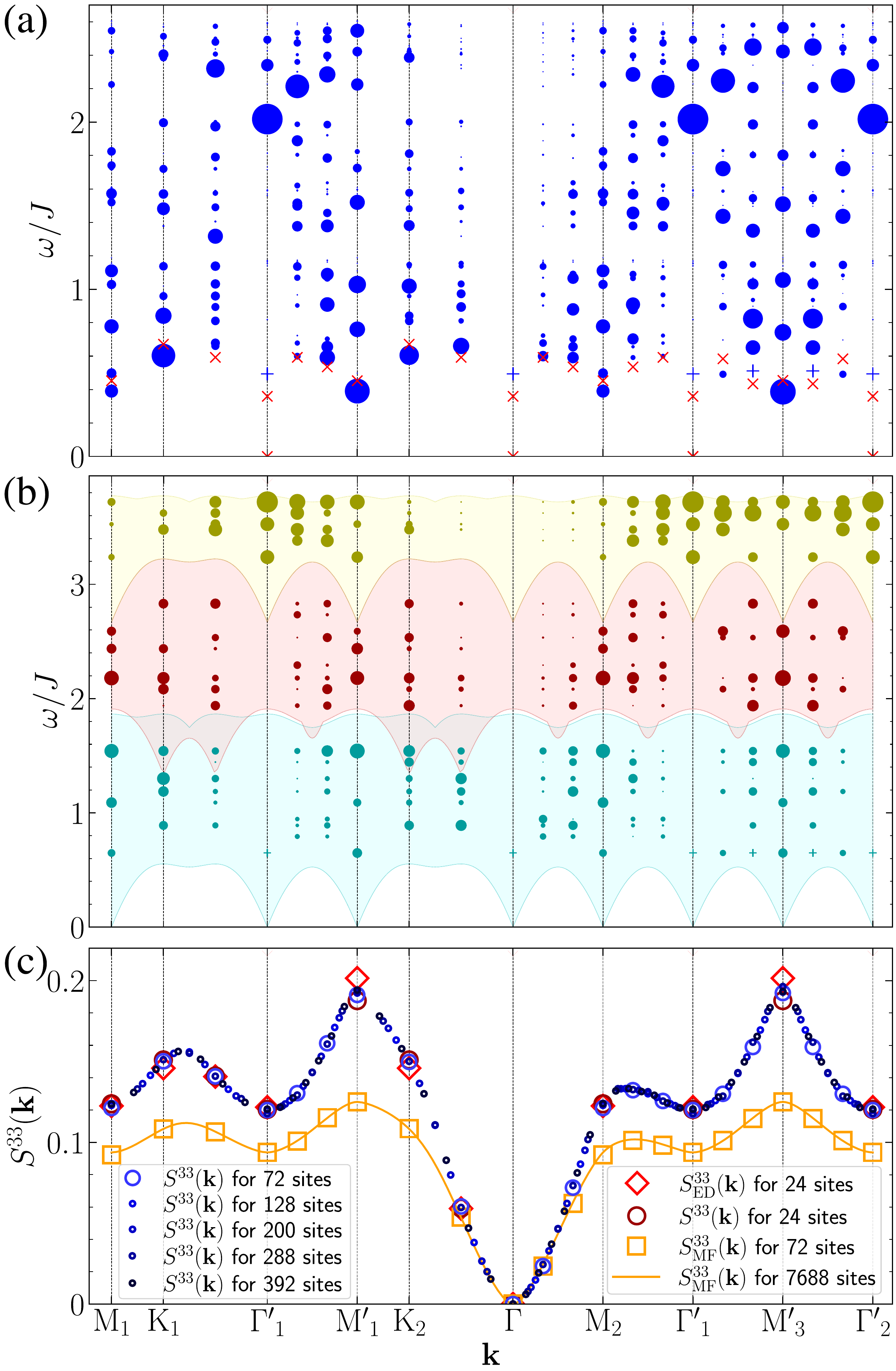}
\caption{
The dynamical structure factor (a) $S^{33}(\mathbf{k},\omega)$ from the variational calculation and (b) $S_{\text{MF}}^{33}(\mathbf{k},\omega)$ [Eq.~(\ref{eq:exact_S^33_for_the_non_interacting_case})] for the mean-field case for a 72-site cluster, along the magenta path in the reciprocal space [see Fig.~\ref{fig:BZ_and_bands}(a)]. 
The area of the circles is proportional to the weights $S^{33}(\mathbf{k},\omega)$.  The color of the circles in (b) corresponds to the colors of the bands of the particle in Fig.~\ref{fig:BZ_and_bands}(b), the filled areas denote the support in the thermodynamic limit. 
The blue `+' denotes the lowest lying excitations of the kind $|\mathbf{k};\mathbf{15}_3; \mathbf{R}, d;\bar{d}\rangle$ with no weight, and the red crosses are the lowest singlets. The mean-field singlet and $\mathbf{15}_3$ particle-hole excitations have equal energies. 
(c) Comparison of the static structure factor $S^{33}(\mathbf{k}) = \int S^{33}(\mathbf{k},\omega) \text{d} \omega$ from the variational and mean-field picture, together with the ED result for the 24-site cluster from  Ref.~\onlinecite{SU4_Honeycomb}. The correlations are smaller in the mean-field case as in the projected case, due to charge fluctuations in $| \pi \text{FS} \rangle$, as explained in Appendix \ref{sec:appendix_sum_rule}. The Monte Carlo errors in (a) and (c) are smaller than the symbol sizes.
\label{fig:S_q_omega_and_S_q_and_spectrum_of_H_MF}
}
\end{figure}

We have compared the $S^{33}(\mathbf{k},\omega)$ calculated by this method and by exact diagonalization for a small 16-site cluster in Fig.~\ref{fig:S_q_omega_16}, where for a well-chosen boundary condition (Fig.~\ref{fig:S_q_omega_16}(c)) the lowest energy excitations seem to be quite similar in the two cases.
Fig.~\ref{fig:S_q_omega_and_S_q_and_spectrum_of_H_MF}(a)  displays the variational $S^{33}(\mathbf{k},\omega)$ for a 72-site cluster. We can recognize towers at low energies centered at the $\text{M}$ and $\text{M}'$ points in the Brillouin zone and at higher energies at the $\Gamma'$ points. 
We will discuss these features in more detail in the next section.

\section{Mean-field dynamical structure factor}

To get an insight into the structure of $S^{33}(\mathbf{k}, \omega)$, we calculated the dynamical structure factor for the mean-field Hamiltonian (\ref{eq:P_G_pi_flux_FS_mean_field_Hamiltonian}). 
Since the $\mathcal{H}_{\text{MF}}$ describes non-interacting fermions, the 
$S_{\text{MF}}^{33}(\mathbf{k}, \omega)$ in Eq.~(\ref{eq:S^33_k_omega}) can be evaluated exactly. The $|\pi \text{FS}\rangle$ is a filled sea of one-particle states, and the excitations correspond to moving particles from the Fermi sea to the unoccupied bands.
Thus moving a flavor-$\alpha$ fermion creates the excitation $c^{\dagger}_{\mathbf{k+q}, n, \alpha} c^{\phantom{\dagger}}_{\mathbf{q}, n', \alpha} |\pi \text{FS} \rangle$ with momentum $\mathbf{k}$, where the band indices are $n'\in \{ 1,2\}$ and $n \in \{ 3,\dots 8 \}$ (the bands are twofold degenerate, in Fig.~{\ref{fig:BZ_and_bands}}(b) the color of the lowest energy band with $n = 1,2$ is green, the second band with $n=3,4$ is cyan, the third one with $n=5,6$ is red, and the highest band with $n=7,8$ is yellow). 
As shown in appendix \ref{sec:appendix_S_q_w_MF_derivation}, using the 
\begin{equation}
 c^{\dagger}_{\mathbf{R}, d, \alpha} =  \sqrt{\frac{8}{N_s}} \sum_{\mathbf{q}\in \text{rBZ}} e^{-i \mathbf{q} \cdot \mathbf{R}} c^{\dagger}_{\mathbf{q}, d, \alpha},
 \end{equation}
we get
\begin{multline}
    S_{\text{MF}}^{33}(\mathbf{k},\omega) 
    = \frac{1}{2N_s} 
    \sum_{\substack{ \mathbf{q} \in \text{rBZ} \\ n \in \{ 3 \dots 8 \} \\ n' \in \{1,2\} } } 
    \left| \sum_{d=A}^{H}     e^{i \mathbf{k} \cdot \boldsymbol{\delta}_d} v^{*}_{\mathbf{k} + \mathbf{q}, n, d} v_{ \mathbf{q}, n, d'} \right|^2  
    \\ \times 
    \delta(\omega - \lambda_{\mathbf{k} + \mathbf{q},n} + \lambda_{ \mathbf{q},n'})\;,
    \label{eq:exact_S^33_for_the_non_interacting_case}
\end{multline}
where $v_{ \mathbf{q}, n, d}$ are the coefficients of the one-particle eigenstate of $\mathcal{H}_{\text{MF}}$ created by
 $c^{\dagger}_{\mathbf{q}, n, \alpha} = \sum_{d = \text{A}}^{\text{H}} v_{ \mathbf{q}, n, d} c^{\dagger}_{\mathbf{q}, d, \alpha}$
 in band with index $n$, with momentum $\mathbf{q}$ and energy $\lambda_{\mathbf{q},n}$.
The index of the basis sites takes eight values $d \in \{ \text{A} ... \text{H}\}$, in accordance with a unit cell of the $\mathcal{H}_{\text{MF}}$ drawn  Fig.~{\ref{fig:lattice}}(b). The reduced Brillouin zone (rBZ) of the eight-site unit cell contains $N_s/8$ momenta and it is the smallest hexagon in Fig.~\ref{fig:BZ_and_bands}(a); it determines the periodicity of the bands and the excitation energies. 
For example, a particle-hole excitation with $\mathbf{k} =0$ in the reduced Brillouin zone describes excitations not only at the $\Gamma$ point in the reciprocal space but also at the $\text{M}$, $\text{M}'$, $\Gamma'$, etc., points. 
The $e^{i \mathbf{k}\cdot \boldsymbol{\delta}_d}$ phases in the matrix elements make the weight of excitation in the $ S_{\text{MF}}^{33}(\mathbf{k},\omega)$ periodic with $\mathbf{Q} = (4\pi,0)$ and $(2\pi,\pm 2\pi \sqrt{3})$, i.e.,  $ S_{\text{MF}}^{33}(\mathbf{k},\omega) =  S_{\text{MF}}^{33}(\mathbf{k}+\mathbf{Q},\omega)$. 
So it is enough to calculate the $ S^{33}(\mathbf{k},\omega) $ and $ S_{\text{MF}}^{33}(\mathbf{k},\omega) $ within the extended Brillouin zone, see Fig.~\ref{fig:BZ_and_bands}(a).
We evaluated Eq.~(\ref{eq:exact_S^33_for_the_non_interacting_case}) numerically for the same cluster used for the variational calculation,
 the resulting $S_{\text{MF}}^{33}(\mathbf{k},\omega)$ is presented in Fig.~\ref{fig:S_q_omega_and_S_q_and_spectrum_of_H_MF}(b).
As explained in Appendix \ref{sec:appendix_sum_rule}, the mean-field sum rule is smaller than in the projected case, Eq.~(\ref{eq:sum_rule_GP}), due to charge fluctuations in $| \pi \text{FS} \rangle$, 
\begin{equation}
  \sum_{\mathbf{k}\in \text{eBZ}}\int S_{\text{MF}}^{33}(\mathbf{k},\omega) \text{d} \omega = \frac{9}{64} N_{s}.
\label{eq:sum_rule_MF}
\end{equation}

\begin{figure}[!h]
\centering
\includegraphics[width=0.47\textwidth]{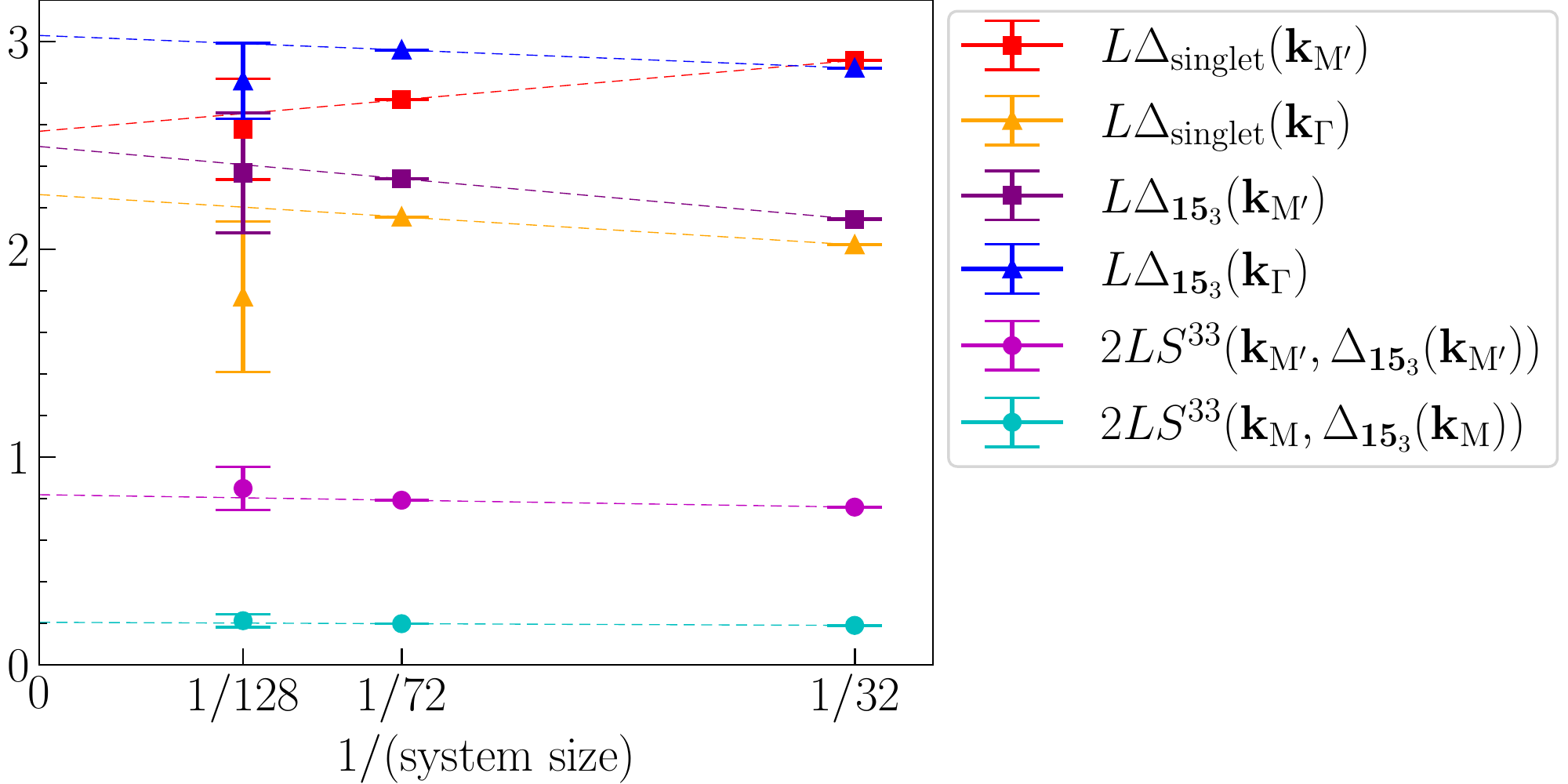}
\caption{
In the variational calculation, the $L \Delta(\mathbf{k})$ at the $\mathrm{M'}$ and $\mathrm{\Gamma}$ points tend to a finite value, indicating that the gap probably vanishes as $\Delta(\mathbf{k}) \propto L^{-1}$, just like in the mean-field case. Similarly, the weights at the bottom of the towers at the $\text{M}'$ and $\text{M}$ points probably scale as $S^{33}(\mathbf{k}, \Delta(\mathbf{k})) \propto L^{-1} $, unlike in the mean-field case. However, the error bars in both cases are too big to draw definitive conclusions.
The straight lines go through the 32- and 72-site results and serve as a guide to the eye. The number of independent samples (separated by a number of elementary steps corresponding to the correlation time) for sizes 32, 72, and 128, were approximately $10^9$,  $10^{8}$, and between $10^{8}$ and $10^{9}$, respectively.
\label{fig:gap_scaling}}
\end{figure}

First, we concentrate on the low energy behavior of the $S^{33}(\mathbf{k},\omega)$ and  $S_{\text{MF}}^{33}(\mathbf{k},\omega)$. 
The lowest energy peaks are at the $\text{M}$ and $\text{M}'$ points in the extended Brillouin zone, as shown in Fig.~\ref{fig:S_q_omega_and_S_q_and_spectrum_of_H_MF}(a) and (b). In the mean-field picture, they originate from the particle-hole excitation $c^{\dagger}_{ \mathbf{q}, n} c^{\phantom{\dagger}}_{\mathbf{q}, n'} |\pi \text{FS} \rangle$ from the top of the Fermi sea  ($n'=1,2$) to the lowest unoccupied states at the bottom of the Dirac cone in the second band ($n=3,4$), highlighted by the arrow labeled $\omega_{\text{M}}$ in Fig.~\ref{fig:BZ_and_bands}(b). 
The precise analysis is complicated by the necessity of the antiperiodic boundary condition for the mean-field problem; the momenta of the fermions are shifted relative to the momenta in the $S^{33}(\mathbf{k},\omega)$. No momentum is available at the $\Gamma$ point, the Dirac point cannot be occupied. Instead, there are two available momenta near the Dirac Fermi point, both for the holes and for the particles (the symmetry of the mean-field problem is the $D_2$, the dashed green line in Fig.~\ref{fig:BZ_and_bands}(a) and (b) represent the reflections). Consequently, instead of a single peak, three peaks appear at the bottom of each tower at the $\text{M}$, $\text{M}'$, and around the $\Gamma$ and $\Gamma'$ points. 
We can see these peaks along the $\Gamma_1'$ -- $\text{M}_3'$ -- $\Gamma_2'$ path in Fig.~\ref{fig:S_q_omega_and_S_q_and_spectrum_of_H_MF}(b) and at the same energy for the $S_{\text{MF}}^{33}(\mathbf{k},\omega)$. These peaks are also present in the variational calculation in Fig.~\ref{fig:S_q_omega_and_S_q_and_spectrum_of_H_MF}(a), but compared to the mean-field case, the weights of the satellite peaks are smaller, and their energies increased. In the thermodynamic limit, the two momenta near the Dirac-Fermi point merge; therefore, it is intuitively clear why the mean-field energy spectrum becomes gapless.
The finite-size scalings of the gaps at the $\text{M}'$ and $\Gamma$ points for the projected case suggest a $\Delta(\mathbf{k}) \propto L^{-1} + \mathcal{O}(L^{-3})$ scaling (though the error bars are large, as shown in Fig.~\ref{fig:gap_scaling}), which would imply a gapless spectrum, just like in the mean-field case. 
The projected spectrum may have a gap if the Gutzwiller projector is able to open the gap. Since the differences between the projected and mean-field calculations become smaller with increasing $N$, if the Gutzwiller projector cannot open a gap in the SU(2) case, we do not expect a gap in the SU(4) case either. While we are not aware of a finite-size scaling analysis of the gap in the SU(2) case, the gapless feature of the projected spectrum is quite convincing in References \onlinecite{Becca,2018PhRvB..98j0405F,2019PhRvX...9c1026F,2020JPCM...32A4003F,Ferrari_kagome}.

%
%
Overall, the energy spectrum is very similar in the two cases. We can think of the towers as the analogs of the two-spinon continuum of the one-dimensional $S=1/2$ Heisenberg model. 
Let us also note that for the staggered-flux SU(2) Dirac spin liquid on the triangular lattice, the Gutzwiller projector gives rise to a  well-pronounced low-energy mode apart from the continuum, which has no counterpart in the mean-field case \cite{2019PhRvX...9c1026F, wietek2023quantum}, reflecting the vanishing role of charge fluctuations in the large-$N$ limit \cite{Affleck_pi_flux, PhysRevB.39.11538}.

The main difference between $S^{33}_{\text{MF}}(\mathbf{k}, \omega)$ and $S^{33}(\mathbf{k}, \omega)$ is in the distribution of the weights. For the projected case, there is a tendency to intensify the lower energy weights, thus making the lower edge of the towers more prominent.
The lowest energy weight of $S^{33}_{\text{MF}}(\mathbf{k}, \omega)$ at the $\text{M}'$ points scales as $\frac{1}{3} L^{-2} + \mathcal{O}\left(L^{-4}\right)$, and at the $\text{M}$ points as $\frac{5}{24} L^{-2} + \mathcal{O}\left({L^{-4}}\right)$, in $N_s=2L^2$ clusters.
However, the scaling of the lowest weight in the projected calculations seems to follow a different scaling, $S^{33}(\mathbf{k}, \Delta(\mathbf{k})) \propto L^{-1} + \mathcal{O}(L^{-3})$, as shown in Fig.~\ref{fig:gap_scaling}. 
Unfortunately, the large error bars do not allow a more precise determination of the finite-size scalings. 
We also found that at low energies, the local $S^{33}_{\text{MF}}(\omega) = \sum_{\mathbf{k}}\int S^{33}_{\text{MF}}(\mathbf{k},\omega)$ within a tower above the $M$ and $M'$ wave vectors is proportional to the degeneracy (the number of 
particle-hole excitations with the same $\omega$), ignoring finite size corrections.  
Consequently, the 
$\int_0^{\omega} S^{33}_{\text{MF}}(\omega') \text{d} \omega' \propto \omega^4$ and from this the $S^{33}_{\text{MF}}(\omega) \propto \omega^3$ follows. However, the 
matrix elements in Eq.~(\ref{eq:exact_S^33_for_the_non_interacting_case})
are not all equal at a given $\omega$.

At the  $\Gamma'$ points, three excitations with non-zero weight appear one above the other at relatively high energy for both the mean-field and the projected cases, shown in Figs.~\ref{fig:S_q_omega_and_S_q_and_spectrum_of_H_MF}(a) and (b).
Again, the projected calculations show a weight shift from higher to lower energies. 
Analyzing the mean-field case, the lowest peak with non-zero weight comes from the particle-hole excitation $c^{\dagger}_{\mathbf{q}, n} c^{\phantom{\dagger}}_{\mathbf{q}, n'} |\pi \text{FS} \rangle$ with a fermion being moved from the top of the Fermi sea ($n' \in \{ 1,2 \}$) to the lowest unoccupied states at the bottom of the Dirac cone in the highest energy band ($n \in \{ 7,8 \}$), indicated by the $\omega_{\Gamma'}$ arrow in Fig.~{\ref{fig:BZ_and_bands}}, so that $\omega_{\Gamma'} \rightarrow 2\sqrt{3} t \approx 2.67 J$ in the thermodynamic limit.

Even though the towers are located at the $\text{M}$ and $\text{M}'$ points, the static spin structure factor 
\begin{align}
  S^{33}(\mathbf{k}) &= \int S^{33}(\mathbf{k}, \omega) \text{d}\omega \nonumber\\
  & = \langle \pi \text{FS} |P_{\text{G}} T^{3}_{-\mathbf{k}} T^{3}_{\mathbf{k}} P_{\text{G}} | \pi \text{FS} \rangle
\end{align}
has maxima only at the $\text{M}'$ points in the form of a cusp.
Physically, we may attribute the cusp to a short-range four-sublattice order \cite{SU4_Honeycomb}. 
We also compared the static structure factors: the projected one with the exact one for the 24-site cluster from Ref.~\onlinecite{SU4_Honeycomb}, and the mean-field and the projected one for larger system sizes in Fig.~\ref{fig:S_q_omega_and_S_q_and_spectrum_of_H_MF}(c). The mean-field static spin correlations of $|\pi\text{FS}\rangle$ decay as $1/\text{distance}^4$, typical behavior for systems with Dirac Fermi points \cite{Affleck_pi_flux,rantner2002}. This leads to a $|\mathbf{k}-\mathbf{k}_{\text{M}'}|^2 \log |\mathbf{k}-\mathbf{k}_{\text{M}'}|$ like weak singularity in $S_{\text{MF}}^{33}(\mathbf{k})$, which is not visible in Fig.~\ref{fig:S_q_omega_and_S_q_and_spectrum_of_H_MF}(c).
For comparison, according to \cite{SU4_Honeycomb}, the static correlations of $P_{\text{G}} |\pi \text{FS} \rangle$ decay with distance with power between $-4$ and $-3$, giving a cusp and a noticeable finite size dependence at the $\text{M}'$ point.

\section{Conclusion}

In conclusion, we calculated the dynamical structure factor $S(\mathbf{k},\omega)$ in the SU(4) Heisenberg model on the honeycomb lattice, assuming the $\pi$-flux Dirac spin liquid variational ground state. 
We find a qualitative agreement between the mean field and the Gutzwiller projected variational treatment, indicating the vanishing importance of charge fluctuations as the number of flavors increases.
 Besides the gapless continuum of excitations originating from the Dirac cones, a similar continuum appears at higher energies in $S(\mathbf{k},\omega)$. The projection's primary effect is shifting the weights to lower energies. In the mean-field calculation, we recovered the algebraic decay of the spin correlations $S_{\text{MF}}(r) \propto 1/r^{4}$ with the distance $r$ expected for a filled Dirac Fermi sea, and obtained the $S_{\text{MF}}(\omega) \propto \omega^3$ behavior for the local dynamical correlations.
 
\begin{acknowledgments}

This work was supported by the Hungarian NKFIH Grant No. K 142652. We acknowledge discussions with Hong-Hao Tu, Yasir Iqbal, and Zoltán Vörös.

\end{acknowledgments}

\appendix

\section{Derivation of $S^{33}_{\text{MF}}(\mathbf{k},\omega)$}
\label{sec:appendix_S_q_w_MF_derivation}

The Bravais lattice vectors of the quadrupled unit cell of the mean-field Hamiltonian $\mathcal{H}_{\text{MF}}$ are $\mathbf{R} = R_{1} 2 \mathbf{a}_1 + R_{2} 2 \mathbf{a}_2$ with $R_{1}, R_{2}$ integers, and the indices of the sites in the basis are $d \in \{ A, \dots, H \}$. 
Substituting
\begin{align}
 T^{3}_{\mathbf{k}} &= \frac{1}{\sqrt{N_s}} \sum_{\mathbf{R},d} e^{i \mathbf{k} \cdot (\mathbf{R}+\boldsymbol{\delta}_d)} T^{3}_{\mathbf{R},d}  \\
 &= \frac{1}{\sqrt{N_s}} \sum_{\mathbf{R},d} e^{i \mathbf{k} \cdot (\mathbf{R}+\boldsymbol{\delta}_d)} \frac{1}{2} \left( c^{\dagger}_{\mathbf{R},d,1} c^{\phantom{\dagger}}_{\mathbf{R},d,1} - c^{\dagger}_{\mathbf{R},d,2} c^{\phantom{\dagger}}_{\mathbf{R},d,2} \right) \nonumber
\end{align}
 into Eq.~(\ref{eq:S^33_k_omega}), we get
\begin{widetext}
\begin{align}
    S^{33}_{\text{MF}}(\mathbf{k},\omega) &= \frac{1}{N_s} \sum_{f} \bigg|  \sum_{\mathbf{R},d} e^{i \mathbf{k} \cdot (\mathbf{R}+\boldsymbol{\delta}_d)} 
     \frac{1}{2} \langle f| c^{\dagger}_{\mathbf{R},d,1} c^{\phantom{\dagger}}_{\mathbf{R},d,1} - c^{\dagger}_{\mathbf{R},d,2} c^{\phantom{\dagger}}_{\mathbf{R},d,2} | \pi \text{FS} \rangle  \bigg|^2
   \delta(\omega + E_{\pi \text{FS}} - E_{f}), \;
\end{align}
where $N_s$ is the number of lattice sites. Using $c^{\dagger}_{\mathbf{R}, d, \alpha} \equiv \frac{1}{\sqrt{N_{\mathbf{C}}}} \sum_{\mathbf{q}} e^{-i \mathbf{q} \cdot \mathbf{R}} c^{\dagger}_{\mathbf{q}, d, \alpha}$ (where $N_{\mathbf{C}} = N_s/8$ is the number of quadrupled unit cells), we obtain
\begin{align}
    S^{33}_{\text{MF}}(\mathbf{k},\omega) &=
    \frac{1}{N_s N_{\mathbf{C}}} \sum_{f} \bigg| \sum_{\substack{ d\\ \mathbf{q},\mathbf{q'}\in \text{rBZ}}} e^{i \mathbf{k} \cdot \boldsymbol{\delta}_d}  \sum_{\mathbf{R}}e^{i (\mathbf{k} - \mathbf{q'} + \mathbf{q})\cdot \mathbf{R}}   
    \frac{1}{2} \langle f|c^{\dagger}_{\mathbf{q'},d,1} c^{\phantom{\dagger}}_{\mathbf{q},d,1} - c^{\dagger}_{\mathbf{q'},d,2} c^{\phantom{\dagger}}_{\mathbf{q},d,2} | \pi \text{FS} \rangle  \bigg|^2 
\delta(\omega + E_{\pi \text{FS}} - E_{f})\;,\;
\end{align}
\end{widetext}
where $\mathbf{q}$ and $\mathbf{q'}$ are wave vectors in the reduced Brillouin zone (rBZ), but $\mathbf{k}$ can be anywhere in the extended Brillouin zone (eBZ), shown in Fig.~\ref{fig:BZ_and_bands}(a). Therefore, we get $\sum_{\mathbf{R}}e^{i (\mathbf{k} - \mathbf{q'} + \mathbf{q})\cdot \mathbf{R}} = N_C \delta_{ \mathbf{q'}, \mathbf{k} + \mathbf{q} + \mathbf{Q}}$, where $\mathbf{Q}$ is the reciprocal lattice vector which maps the wave vector $\mathbf{k} + \mathbf{q} \in \text{eBZ}$ back to the reduced Brillouin zone. Thus, we will hide the wave vector $\mathbf{Q}$ and write everywhere $\mathbf{k} + \mathbf{q}$, which is meant to be mapped back into the reduced Brillouin zone. Consequently, 
\begin{widetext}
\begin{align}
    S^{33}_{\text{MF}}(\mathbf{k},\omega) &=
    \frac{1}{4 N_s} \sum_{f} \bigg| \sum_{d} e^{i \mathbf{k} \cdot \boldsymbol{\delta}_d} \sum_{\mathbf{q}} 
    \langle f|c^{\dagger}_{\mathbf{\mathbf{k} + \mathbf{q}},d,1} c^{\phantom{\dagger}}_{\mathbf{q},d,1} - c^{\dagger}_{\mathbf{\mathbf{k} + \mathbf{q}},d,2} c^{\phantom{\dagger}}_{\mathbf{q},d,2} | \pi \text{FS} \rangle  \bigg|^2  
     \delta(\omega + E_{\pi \text{FS}} - E_{f})\;.
\label{eq:S_MF_with_sum_Q}
\end{align}
The one-particle eigenstate of $\mathcal{H}_{\text{MF}}$ in band $n$ with momentum $\mathbf{q}$ and energy $\lambda_{\mathbf{q},n}$ can be written as a linear combination $c^{\dagger}_{\mathbf{q}, n, \alpha}|0 \rangle = \sum_{d = \text{A}}^{\text{H}} v_{ \mathbf{q}, n, d} c^{\dagger}_{\mathbf{q}, d, \alpha}|0 \rangle $, where $|0 \rangle$ is the vacuum. Inverting this relation we have $c^{\dagger}_{\mathbf{q}, d, \alpha}|0 \rangle = \sum_{n = 1}^{8} v^{-1}_{ \mathbf{q}, d, n} c^{\dagger}_{\mathbf{q}, n, \alpha}|0 \rangle = \sum_{n = 1}^{8} v^{*}_{ \mathbf{q}, n, d} c^{\dagger}_{\mathbf{q}, n, \alpha}|0 \rangle $, where we used that the matrix composed of the coefficients $v_{ \mathbf{q}, n, d}$ is unitary, namely $v^{-1}_{ \mathbf{q}, d, n} = v^{*}_{ \mathbf{q}, n, d}$. 
From these relations, we get
\begin{align}
    c^{\dagger}_{\mathbf{k} + \mathbf{q}, d, \alpha} c^{\phantom{\dagger}}_{\mathbf{q},d,\alpha} |\pi \text{FS} \rangle &= \left( \sum_{n = 1}^{8} v^{*}_{ \mathbf{k} + \mathbf{q}, n, d} c^{\dagger}_{\mathbf{k} + \mathbf{q} , n, \alpha} \right)  
    \left( \sum_{n' = 1}^{8} v_{ \mathbf{q}, n, d'} c^{\phantom{\dagger}}_{\mathbf{q}, n', \alpha} \right) |\pi \text{FS} \rangle \nonumber.
\end{align}
Therefore, we can write 
\begin{align}
    S^{33}_{\text{MF}}(\mathbf{k},\omega) &=
    \frac{1}{4 N_s} \sum_{f} \bigg| \sum_{d} e^{i \mathbf{k} \cdot \boldsymbol{\delta}_d} 
    \sum_{\mathbf{q}} \sum_{n,n'} v^{*}_{ \mathbf{k} + \mathbf{q} , n, d} v_{ \mathbf{q}, n, d'} 
\langle f |  c^{\dagger}_{\mathbf{k} + \mathbf{q}, n, 1}  c^{\phantom{\dagger}}_{\mathbf{q}, n', 1} - c^{\dagger}_{\mathbf{k} + \mathbf{q}, n, 2}  c^{\phantom{\dagger}}_{\mathbf{q}, n', 2}  |\pi \text{FS} \rangle
    \bigg|^2 
    \delta(\omega + E_{\pi \text{FS}} - E_{f})\;.
\end{align}
\end{widetext}
The states $|1 \rangle = c^{\dagger}_{\mathbf{k} + \mathbf{q}, n, 1}  c^{\phantom{\dagger}}_{\mathbf{q}, n', 1}  |\pi \text{FS} \rangle$ and $|2 \rangle = c^{\dagger}_{\mathbf{k} + \mathbf{q}, n, 2}  c^{\phantom{\dagger}}_{\mathbf{q}, n', 2}  |\pi \text{FS} \rangle$ are multiparticle eigenstates of $\mathcal{H}_{\text{MF}}$ having the same excitation energy $ E_f - E_{\pi \text{FS}} = \lambda_{\mathbf{k} + \mathbf{q},n} - \lambda_{\mathbf{q},n'}$. 
For $\mathbf{k} \neq \mathbf{0}$ these states are orthogonal because they represent particle-hole excitation of different flavors. 
However, for $\mathbf{k} = \mathbf{0}$ they are equal, since then $|\alpha \rangle = c^{\dagger}_{\mathbf{q}, n, \alpha}  c^{\phantom{\dagger}}_{\mathbf{q}, n', \alpha}  |\pi \text{FS} \rangle = \delta_{n,n'} \text{n}_{\mathbf{q}, n', \alpha}|\pi \text{FS} \rangle  = \delta_{n,n'} |\pi \text{FS} \rangle$ for both $\alpha \in \{ 1 , 2 \}$, where $\text{n}_{\mathbf{q}, n', \alpha}$ is the density of flavor $\alpha$.   Consequently, $S^{33}_{\text{MF}}(\mathbf{k} = \mathbf{0},\omega) = 0$, reflecting the singlet nature of the Fermi sea. 
%
The $c^{\dagger}_{\mathbf{k} + \mathbf{q}, n, \alpha} c^{\phantom{\dagger}}_{\mathbf{q}, n', \alpha}|\pi \text{FS} \rangle $ may be finite only when $n \in \{ 3 \dots  8 \}$ and $n' \in \{ 1,2 \}$, because only the two-fold degenerate lowest energy band fills the Fermi sea, $|\pi \text{FS} \rangle = \Pi_{\mathbf{q}} \Pi_{n = 1}^{2}\Pi_{\alpha = 1}^{4} c^{\dagger}_{\mathbf{\mathbf{q}}, n, \alpha} |0 \rangle$. 
Therefore, we can restrict the values of $n$ and $n'$ in $\sum_{n,n'}$ to $\sum_{n = 3}^8 \sum_{n'=1}^2 $.
The states $|f \rangle$ giving non-zero value are of the form $|f \rangle = | \mathbf{k}, \tilde{\mathbf{q}}, \tilde{n}, \tilde{n}', \tilde{\alpha} \rangle = c^{\dagger}_{\mathbf{k} + \tilde{\mathbf{q}}, \tilde{n}, \tilde{\alpha}}  c^{\phantom{\dagger}}_{\tilde{\mathbf{q}}, \tilde{n}', \tilde{\alpha}}  |\pi \text{FS} \rangle$, for which we get
\begin{multline}
    \langle f | \left( c^{\dagger}_{\mathbf{k} + \mathbf{q}, n, 1}  c^{\phantom{\dagger}}_{\mathbf{q}, n', 1} - c^{\dagger}_{\mathbf{k} + \mathbf{q}, n, 2}  c^{\phantom{\dagger}}_{\mathbf{q}, n', 2} \right) |\pi \text{FS} \rangle  \\
     = (1- \delta_{\mathbf{k},\mathbf{0}}) \delta_{\tilde{\mathbf{q}},\mathbf{q}} \delta_{\tilde{n},n} \delta_{\tilde{n}',n'} 
    (\delta_{\tilde{\alpha}, 1} - \delta_{\tilde{\alpha}, 2}).
    \label{eq:braket}
\end{multline}
The sum $\sum_{f}$ can be replaced by $\sum_{\tilde{\mathbf{q}} } \sum_{\tilde{n} = 3}^{8}  \sum_{\tilde{n}' = 1}^{2} \sum_{\tilde{\alpha} = 1,2}$, and the sums $\sum_{\mathbf{q}} \sum_{n = 3}^{8}  \sum_{n' = 1}^{2}$ in the absolute value squared disappears due to $\delta_{\tilde{\mathbf{q}},\mathbf{q}} \delta_{\tilde{n},n} \delta_{\tilde{n}',n'}$ in Eq.~(\ref{eq:braket}).  The sum over $\sum_{\tilde{\alpha} = 1}^2$ gives a factor of 2 since the absolute value squared is the same for both $\tilde{\alpha} = 1, 2$. Eventually, the dynamical structure factor in the mean-field approach reads
\begin{align}
    S^{33}_{\text{MF}}(\mathbf{k},\omega) =
    \frac{1}{ 2N_s} \sum_{\tilde{\mathbf{q}} } &\sum_{\tilde{n} = 3}^{8}  \sum_{\tilde{n}' = 1}^{2} \left| \sum_{d} e^{i \mathbf{k} \cdot \boldsymbol{\delta}_d} 
    v^{*}_{ \mathbf{k} + \tilde{\mathbf{q}} , \tilde{n}, d} v_{ \tilde{\mathbf{q}}, \tilde{n}', d} 
    \right|^2 \nonumber \\
     &\times \delta(\omega  - \lambda_{\mathbf{k} + \tilde{\mathbf{q}},\tilde{n}} + \lambda_{\tilde{\mathbf{q}},\tilde{n}'})\;,
\end{align}
where we can leave off the $\tilde{ }$ notation for convenience. 

\section{Sum rules}
\label{sec:appendix_sum_rule}

As we stated in Eqs.~(\ref{eq:sum_rule_GP}) and (\ref{eq:sum_rule_MF}), the sum rules are different in the projected and the mean-field cases.
This section shows how the charge fluctuations affect the sum rule for the mean-field case for a general SU($N$) model in the fundamental representation.
The sum rule is defined as
\begin{align}
    &\sum_{\mathbf{k} \in \text{eBZ}} S^{33}(\mathbf{k}) = \sum_{\mathbf{k} \in \text{eBZ}} \langle \text{GS} | T^{3}_{-\mathbf{k}} T^{3}_{\mathbf{k}}  | \text{GS} \rangle \\
    &= \sum_{\mathbf{k} \in \text{eBZ}} \frac{1}{N_s} \sum_{\substack{ \mathbf{R}, d \\ \bar{\mathbf{R}}, \bar{d}}} e^{i \mathbf{k} \cdot (\mathbf{R} + \boldsymbol{\delta}_d - \bar{\mathbf{R}} - \boldsymbol{\delta}_{\bar{d}})} \langle \text{GS} | T^{3}_{\mathbf{R},d} T^{3}_{\bar{\mathbf{R}},\bar{d}}  | \text{GS} \rangle . \nonumber 
\end{align}
Using the relation
\begin{equation}
    \sum_{\mathbf{k} \in \text{eBZ}} e^{i \mathbf{k} \cdot (\mathbf{R} + \boldsymbol{\delta}_d - \bar{\mathbf{R}} - \boldsymbol{\delta}_{\bar{d}})} = N_{\mathbf{k}} \delta_{\mathbf{R},\bar{\mathbf{R}}} \delta_{d, \bar{d}} ,
\end{equation}
where $N_{\mathbf{k}} = \frac{3}{2} N_s$ is the number of wave vectors in the extended Brillouin zone, we get
\begin{align}
    \sum_{\mathbf{k} \in \text{eBZ}} S^{33}(\mathbf{k}) &= \frac{3}{2} \sum_{ \mathbf{R}, d} \langle \text{GS} | T^{3}_{\mathbf{R},d} T^{3}_{\mathbf{R},d}  | \text{GS} \rangle \\
    &= \frac{3}{2} \sum_{ \mathbf{R}, d} \frac{1}{N^2 - 1} \sum_{a=1}^{N^2 - 1} \langle \text{GS} | T^{a}_{\mathbf{R},d} T^{a}_{\mathbf{R},d}  | \text{GS} \rangle ,\nonumber
\end{align}
where in the last step we assume that the ground state does not break the SU($N$) symmetry (i.e., it is an SU($N$) singlet). The sum rule is proportional to the value of the quadratic Casimir operator, defined as 
\begin{equation}
 \hat C_2 = \sum_{a=1}^{N^2 - 1} T^{a} T^{a} 
\label{eq:Casimir}
\end{equation}
The Casimir operator is diagonal in an irreducible representation, $\hat C_2 = C_2 \mathbf{I}$, where the identity matrix $\mathbf{I}$ has the dimension of the irreducible representation, which is $N$ in the fundamental one. We can get the value of $C_2$ by taking the trace $\text{Tr} \, \hat C_2 = C_2 \text{Tr}\, \mathbf{I} = C_2 N$. Furthermore, using the normalization of spin operators \cite{SU_N_relations}
\begin{equation}
    \text{Tr}\,T^a T^b = \frac{1}{2} \delta_{a,b},
\end{equation}
we get that $\text{Tr} \, \hat C_2 = \sum_{a=1}^{N^2 - 1} \text{Tr} \, T^a T^a = (N^2 - 1)/2$. Thus, in the fundamental representation of SU($N$) the Casimir takes the value
\begin{equation}
  C_2=\frac{N^2 - 1}{2N}.
\end{equation}
The sum rule then reads
\begin{equation}
    \sum_{\mathbf{k} \in \text{eBZ}} S^{33}(\mathbf{k}) = \frac{3}{2} \sum_{ \mathbf{R}, d} \frac{1}{N^2 - 1} C_2 = \frac{3 N_s}{4N},
\end{equation}
resulting in Eq.~(\ref{eq:sum_rule_GP}) for $N=4$.

In the case of the mean-field hopping Hamiltonian of Eq.~(\ref{eq:P_G_pi_flux_FS_mean_field_Hamiltonian}), the value of the quadratic Casimir operator is reduced due to the charge fluctuations in $| \pi \text{FS} \rangle $. To calculate its precise value let us substitute the fermionic parton representation of the SU($N$) spin operators (Eq.~(\ref{eq:spin_operator_with_partons_SU_4}) for $N=4$)
into the Casimir operator
\begin{equation}
    \sum_{a=1}^{N^2 - 1}  T^{a} T^{a} = \frac{1}{4} \sum_{a=1}^{N^2 - 1} \sum_{\substack{\alpha, \beta, \\
    \gamma, \epsilon = 1}}^N c^{\dagger}_{\alpha} \lambda^{a}_{\alpha, \beta} c^{\phantom{\dagger}}_{\beta} c^{\dagger}_{\gamma} \lambda^{a}_{\gamma, \epsilon} c^{\phantom{\dagger}}_{\epsilon}
\end{equation}
where we omit the site indices for convenience. Using the relation $\sum_{a=1}^{N^2 - 1} \lambda^a_{\alpha, \beta} \lambda^a_{\gamma, \epsilon} = 2 \delta_{\alpha, \epsilon} \delta_{\beta, \gamma} - \frac{2}{N} \delta_{\alpha, \beta } \delta_{\gamma, \epsilon}$ \cite{SU_N_relations} we get
\begin{equation}
    \sum_{a=1}^{N^2 - 1}  T^{a} T^{a} = \sum_{\alpha, \gamma = 1}^N \left( \frac{1}{2} c^{\dagger}_{\alpha} c^{\phantom{\dagger}}_{\gamma} c^{\dagger}_{\gamma} c^{\phantom{\dagger}}_{\alpha} - \frac{1}{2N} c^{\dagger}_{\alpha} c^{\phantom{\dagger}}_{\alpha} c^{\dagger}_{\gamma} c^{\phantom{\dagger}}_{\gamma} \right).
\end{equation}
Rearranging the order of the fermionic operators results in
\begin{equation}
    \sum_{a=1}^{N^2 - 1}  T_{\mathbf{R},d}^{a} T_{\mathbf{R},d}^{a} = \frac{N+1}{2} \text{n}_{\mathbf{R},d} - \frac{N+1}{2N} \text{n}_{\mathbf{R},d}^2,
\end{equation}
where $\text{n}_{\mathbf{R},d} \equiv \sum_{\alpha = 1}^N c^{\dagger}_{\mathbf{R},d,\alpha} c^{\phantom{\dagger}}_{\mathbf{R},d,\alpha}$ is the fermionic density operator on the site $\mathbf{r} = \mathbf{R} +  \boldsymbol{\delta}_{d}$.
For a singly occupied site ($\text{n}_{\mathbf{R},d} = 1$), we get $\sum_{a=1}^{N^2 - 1}  T_{\mathbf{R},d}^{a} T_{\mathbf{R},d}^{a} = \frac{N^2 - 1}{2N}$, implying that a singly occupied site represents a spin correctly (strictly speaking, one would also check the higher order Casimir operators).
However, we get different values for the Casimir operator for other occupation numbers. To calculate the expectation value $\sum_{a=1}^{N^2 - 1} \langle \pi \text{FS}| T_{\mathbf{R},d}^{a} T_{\mathbf{R},d}^{a} | \pi \text{FS} \rangle$ we have to consider all the possible occupations with their probabilities. Since the fermions are uncorrelated in the mean-field approach (i.e., each flavor of fermions occupies a site independently of the other flavors), the probability that $r$ fermions occupy a site is determined by the binomial distribution $P_{\text{binom}}(r) = \binom{N}{r} p^r (1-p)^{N-r}$, where $p=1/N$ is the probability that one of the flavors occupies this site. Thus 
\begin{align}
    \sum_{a=1}^{N^2 - 1} \langle \pi \text{FS}| &T_{\mathbf{R},d}^{a} T_{\mathbf{R},d}^{a} | \pi \text{FS} \rangle   \nonumber \\ 
   & = \sum_{r = 0}^{N}  P_{\text{binom}}(r)   \left( \frac{N+1}{2} r - \frac{N+1}{2N} r^2  \right) \\
    &= \left( \frac{N^2 - 1}{2N}\right) \left(1 - \frac{1}{N} \right),
\end{align}
for any site. The above equation shows that the deviation from the proper value of the quadratic Casimir operator in the fundamental representation is a factor $1 - \frac{1}{N}$. We note that this is an average, as the probability of multiply occupied sites ($r > 1$) is non-zero even in the large $N$ limit, since the binomial distribution $P_{\text{binom}}(r)$ approaches the Poisson distribution with parameter $\lambda = N p = 1$. For $N=4$, the expectation value of the Casimir operator is $45/32$. Therefore, the sum rule is
\begin{equation}
    \sum_{\mathbf{k} \in \text{eBZ}} S^{33}_{\text{MF}}(\mathbf{k}) = \frac{3}{2} \sum_{ \mathbf{R}, d} \frac{1}{15} \frac{45}{32} = \frac{9 }{64}N_s ,
\end{equation}
as written in Eq.~(\ref{eq:sum_rule_MF}).

\section{Projective symmetry in case of antiperiodic boundary condition}
\label{sec:appendix_APBC}

\begin{figure}[ht!]
\includegraphics[width=0.62\columnwidth]{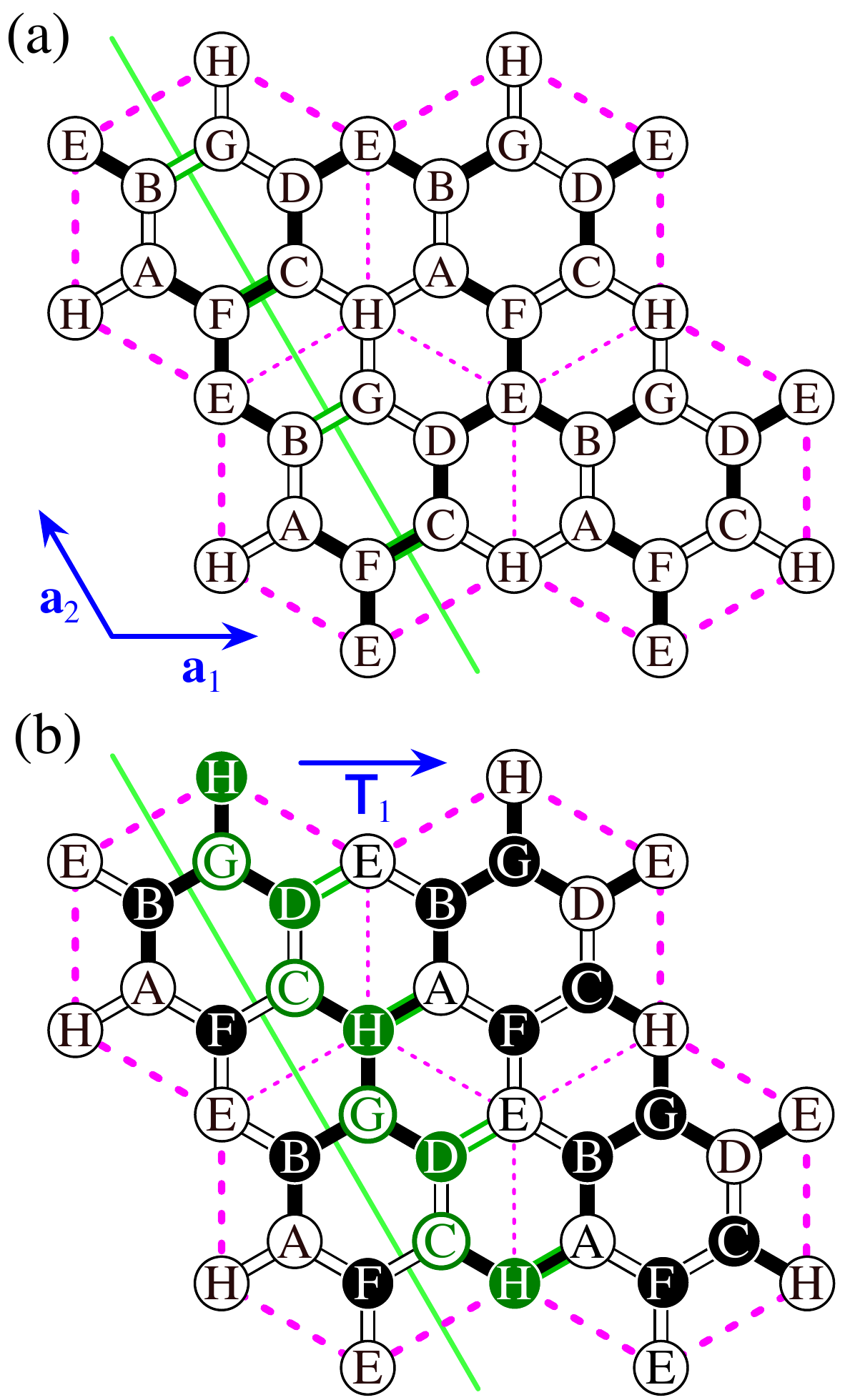}
\caption{
(a) A 32-site cluster containing four quadrupled unit cells shown in Fig.~\ref{fig:lattice}(b), where the black and white bonds represent hopping amplitudes of opposite sign. 
The green line shows the antiperiodic boundary, which flips the sign of the hoppings $t_{i,j}$ crossing it (highlighted in green). 
(b) The translation $\mathsf{T}_1$ by the primitive vector $\mathbf{a}_1$ acts on the fermionic operators as $\mathsf{T}_1 c^{\dagger}_{i} = c^{\dagger}_{\mathsf{T}_1(i)}$ and all the hoppings are shifted to the right, in the direction of $\mathbf{a}_1$. 
For periodic boundaries, the gauge transformation $G^{\text{PBC}}_{\mathsf{T}_{1}}$, which multiplies by $-1$ the fermionic operators on the sublattices B, C, F, and G, restores the original hopping configuration, according to Fig.~\ref{fig:lattice}(g) and Eq.~(\ref{eq:projective_symmetry_for_hoppings}).
However, for an antiperiodic boundary, we must consider the hoppings crossing the green line that were flipped in (a) and shifted to the right in (b). 
Combining the $G^{\text{PBC}}_{\mathsf{T}_{1}}$ with the sign reversal of the fermionic operators highlighted by green along the boundary we get the gauge transformation $G^{\text{APBC}}_{\mathsf{T}_{1}}$. The sign reversal of the fermionic operators marked with solid circles restores the initial hopping configuration of (a) and provides the phases in Eq.~(\ref{eq:projective_symmetry_for_hoppings}) for the antiperiodic case.
}
\label{fig:projective_T1_APBC}
\end{figure}

We impose antiperiodic boundary conditions as shown in Fig.~\ref{fig:lattice}(c) and Fig.~\ref{fig:projective_T1_APBC}(a), which changes the sign of the hoppings crossing one of the cluster's boundaries. 
After these sign changes Eq.~(\ref{eq:projective_symmetry_for_hoppings}) will still hold for
the $\mathsf{T}_2$ translations and the two reflections about the antiperiodic boundary and perpendicular to it (the two reflections generate the $D_2$ point group, which also includes a $\mathsf{C}_2$ rotation), but it will no longer hold for the $\mathsf{T}_1$, $\mathsf{C}_6$ and the remaining reflections if we use the same gauge transformations as we used in the case of periodic boundary conditions.
However, we can satisfy Eq.~(\ref{eq:projective_symmetry_for_hoppings}) for $\mathsf{T}_1$ if we multiply by $-1$ the gauge phases $e^{i \phi_{\mathsf{T}_1}(j)}$ at the locations marked by green circles in Fig.~\ref{fig:projective_T1_APBC}(d). Regarding the $\mathsf{C}_6$ rotations and the reflections about the other axis, we could not find a way to restore the projective symmetry. Indeed, the numerical calculation of the structure factor confirms the $D_2$ point group symmetry for antiperiodic boundary conditions, providing further support for the absence of gauge transformations that would restore the full $D_6$ point group.

Note that the boundary conditions of the fermionic operators do not affect the periodic boundary condition of the spin operators, as can be understood from Eq.~(\ref{eq:spin_operator_with_partons_SU_4}).

\clearpage

\bibliography{su3}

\end{document}